\documentstyle[11pt,psfig,paspconf]{article}
\begin{document}
%\begin{center}
\title{\bf Small scale substructure, collapse time and dynamical friction } 
%\large
\author{M. Gambera} 
\affil{Istituto di Astronomia dell'Universit\`a di Catania, 
Citt\`a Universitaria, Viale A.Doria, 6 - I 95125 Catania, Italy}

\author{A. Pagliaro}
\affil{Istituto di Astronomia dell'Universit\`a di Catania, 
Citt\`a Universitaria, Viale A.Doria, 6 - I 95125 Catania, Italy}

\begin{abstract}
We solve numerically the equations of motion for the collapse 
of a shell of baryonic matter, made of galaxies and substructure 
of $ 10^{6} M_{\odot} {\scriptstyle \div} 10^{9} M_{\odot}$, 
taking into account the dynamical 
friction and the parameters on which it depends: 
the peaks' height $\nu_{c}$, the number of peaks inside a protocluster 
$N_{tot}$ multiplied by the correlation function $\xi (r)$ evaluated 
at $r=0$, the filtering radius $R_{f}$ and  
the core radius of the cluster of galaxies, $r_0$.
We show how the collapse 
time of a shell of baryonic matter depends on those parameters.\\
\end{abstract}
~\\
%\end{abstract}
~\\
~\\
%\begin{center}
%{\bf To be published:}\\
%{\it Proceedings} 
%{\sc Astronomical Society of the Pacific}\\
%Conference Series\\
%book of the International Workshop}\\
%{\sc DARK AND VISIBLE MATTER IN GALAXIES AND COSMOLOGICAL IMPLICATIONS}\\
%{\it Sesto Pusteria - (BZ) 2-5 July 1996\\}
%\end{center}
%\newpage
%\begin{flushleft}
\section{Introduction}
%\end{flushleft} 
%\noindent
The problem of the formation and evolution of clusters of galaxies has been one of the
crucial topic over the last years (see {\it e.g.}  Colafrancesco et al. 1989, 
Kaiser 1993, Antonuccio-Delogu 1992, Colafrancesco et al. 1995 and
Ryden \& Gunn 1987). 
It is well known that the formation of cosmic structures is strictly related to 
the evolution of the density 
perturbations:
in the present {\it paradigma} of structure formation,  it is generally
assumed that cosmic structures 
of size $\sim$ R form preferentially around the local maxima of the 
primordial density field, once it is smoothed on the filtering scale 
$R_{f}$. These linear density fluctuations eventually evolve towards the 
nonlinear regime under the action of gravitational instability; they detach
from the Hubble flow at {\it turn around} epoch $ t_{m}$, given by: 
\begin{equation}
t_{m} = \left[\frac{ 3 \pi}{32 G \rho_{b}} ( 1 +\overline{\delta})
\right]^{1/2} (1+z)^{3/2}
\end{equation}
where $ \rho_{b} $ is the mean background density, $z$ is the redshift and
$\overline{\delta}$ is the mean overdensity within the nonlinear region. After the
 {\it turn around} epoch, the fluctuations  
start to recollapse when their overdensity  reaches the value
$\overline{\delta}=1$. 
Since the density field 
%depend on the  {\sc PS},  which in turn 
depends 
on the matter that dominates the universe, the mean characteristics of the 
cosmic structures depend on the assumed model. In this context the most 
succesful model is the biased Cold Dark Matter  (hereafter {\sc
CDM}) (see {\it e.g.} Liddle $\&$ Lyth 1993)
based on a scale invariant spectrum of density fluctuations growing under 
gravitational instability. In such scenario the formation of the structures occurs 
through a  "{\it bottom up}" mechanism.\\
%Point {\bf (d)}   
A simple model that describes the collapse of a perturbation of density is 
that of Gunn \& Gott (1972, hereafther GG72).
This is in 
contradiction with the predictions of {\sc CDM} models. 
It is well known that in a {\sc CDM} Universe, an 
abundant production of substructures during the evolution of the fluctuations
is predicted.
The presence of substructure is very important for the 
dynamics of collapsing shells of baryonic matter made of galaxies and 
substructure of $ 10^{6} M_{\odot} {\scriptstyle \div} 
10^{9} M_{\odot}$, falling into 
the central regions of a cluster of galaxies. 
In presence of substructure it is necessary 
to modify the equation of motion:
\begin{equation}
\frac{d^{2} r} {dt^{2}} = - \frac{GM}{r^{2} (t)}  
\end{equation} 
since the graininess 
of mass distribution in the system induces dynamical 
friction that introduces 
a frictional force term. 
In a material 
system, the gravitational field can be decomposed into an average field, 
$ {\bf F}_{0}(r)$, generated from the smoothed out distribution 
of mass, and a stochastic component of the field, $ {\bf F}_{stoch}(r)$, generated 
from the fluctuations in number of the neighbouring particles. 
The stochastic component of the gravitational field is 
specified assigning a probability density, $ W( {\bf F})$, (Chandrasekhar \& 
von Neumann 1942, hereafter {\sc CvN42}). In an infinite homogeneous unclustered system 
$ W( {\bf F})$ is given by the Holtsmark distribution  
({\sc CvN42}) whilst in inhomogeneous and clustered systems $ W({\bf F})$ 
is given by Kandrup (1980) and Antonuccio-Delogu \& Atrio-Barandela (1992, hereafter {\sc AA92}) and
respectively. 
The stochastic force, $ {\bf F}_{stoch}$, in a self-gravitating 
system modifies the motion of particles as if it was a frictional force.   
In fact, a particle moving faster than its neighbours produces a deflection 
of their orbits in such a way that the average density is greater in 
the direction opposite to that of motion, causing a slowing down 
in its motion. 
Adopting the notation of {\sc GG72} (see also their eqs. 6 and 8) and 
remembering that $T_{c0}/2$ is the collapse time in
the absence of dynamical friction ({\sc GG72}), one can write:
\begin{equation}
T_{c0} = \frac{\pi \bar{\rho_i} \rho_{ci}^{1/2} } {H_i ( \bar{\rho_i} - \rho_{ci} )^{3/2} }
\end{equation}
where $ \rho_{ci} $ is the 
critical density at a time $ t_{i} $ and $ \bar{\rho_i} $ is 
the {\it average density} inside $ r_{i}$ at $ t_{i}$.
The equation of motion of a shell of baryonic matter in presence of 
dynamical friction, using the dimensionless time variable  
$ \tau = \frac{t}{T_{c0}} $,  can be written in the form:
\begin{equation}
\frac{d^{2} a}{d \tau^{2} }= -
\frac{4 \pi G \rho_{ci}( 1+\overline{\delta_{i}})}{a^{2}(t)} T_{co}^{2}-
%\eta_{0} T_{c0} \frac{ d a }{ d \tau} ( 1+ \frac{\eta}{\eta_{cl}}) 
\eta T_{c0} \frac{ d a }{ d \tau}  
\label{eq:p}
\end{equation}
(Antonuccio-Delogu \& Colafrancesco 1994, hereafter {\sc AC94})
where $ \overline{ \delta_{i}}$ is the overdensity within $ r_{i}$ and
$ \eta$ is the coefficient of dynamical friction. Supposing that there 
are no correlations among random force and their derivatives, we have: 
\begin{equation}
\eta = \frac{\int d^{3} F W(F) F^{2} T(F) }{ 2 \langle v^{2} \rangle }
\label{eq:q}
\end{equation}    
(Kandrup 1980), where $ T(F)$ is the average {\it "duration"} of a random 
force impulse of magnitude $ F$, $ W(F)$ is the probability distribution 
of stochastic force (which for a clustered system is given in eq. 37 of
{\sc AA92}).
~\\
~\\

%\begin{flushleft}
\section{The collapse time} 
%\end{flushleft} 

In a previous paper (Del Popolo \& Gambera 1996, hereafter {\sc DPG96}),
one of us 
showed how the expansion parameter $ a(\tau)$ depends on the dynamical
friction, solving eq.(\ref{eq:p}) numerically. 
{\sc DPG96} assigned some values to $ \eta$ and then
solved eq.(\ref{eq:p}) through a numerical method  
not taking into account  
the parameters on which $ \eta$ depends.
In this paper, we examine how the dynamical friction  coefficient
$ \eta$ varies according to the parameters and how the collapse
time depends on them. 
We solved eq.(\ref{eq:q}) numerically and the other equations 
on which it depends  for a outskirts shell of baryonic matter with
$ \overline{\delta} = 0.01$ inside the spherical regions (protocluster),
for different values of $ \nu_{\it c}$,  $ R_{\it f}$,  $ r_{\it 0}$
and  $ \Xi$, where:
\begin{figure}[ht]
\psfig{file=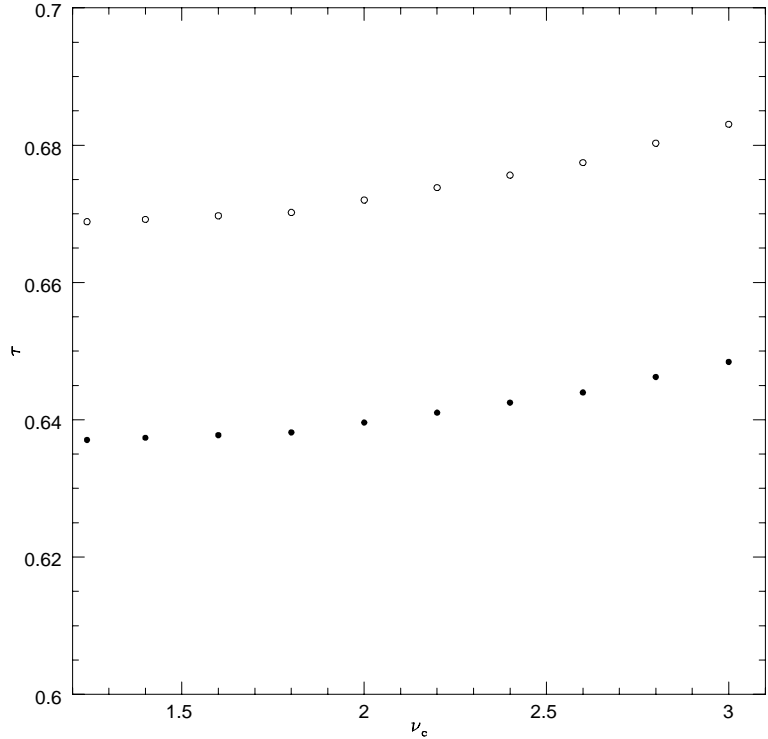,width=11cm}
\caption[h]{ Collapse time $ \tau$
of a shell of matter made of galaxies and substructure 
when dynamical friction is taken into account, 
versus $ \nu_{\it c}$
We assume a core radius of $r_{0} = 1  h^{-1} Mpc $ 
%a central overdensity $\overline{ \delta} =0.01$ 
and a filtering radius $ R_{f} = 0.74 h^{-1} Mpc$.
{\it Open circles:} $ \Xi = 10^{3}$; {\it filled circles:} $ \Xi 
= 10^{2}$.}
\end{figure}
\begin{itemize}
\item $ \; \nu_c$ is the peaks' height;
\item $ \; R_f$ is the filtering radius;
\item $\; r_0$  is the parameter of the power-law density profile;
Theoretical work (Ryden 1988) suggests that the density profile inside a protogalactic dark matter
halo, before relaxation and baryonic infall, can be approximated by a power-law:
\begin{equation}
\rho(r) = \frac{ \rho_0 r_0^p} {r^p}
\end{equation}
where $p \approx 1.6$ on a protogalactic scale. 
\item $\; \Xi \,$ is the product $N_{tot} \cdot \xi (0)$ where $N_{tot}$ is the
total number of peaks inside a protocluster and $\xi(0)$ is the correlation function
calculated in $r=0$.
{\sc AA92} have demonstrated that in the hypothesis $m_{av} \gg 1 M_{\odot} $, 
where $m_{av}$ is the 
average mass of the subpeaks, the dependence of the 
dynamical friction coefficient on $N_{tot}$ and $\xi(r)$ may be  expressed as a dependence on a
single parameter that we define as:
\begin{equation}
\Xi \equiv N_{tot} \cdot  \xi(0)
\end{equation}
\end{itemize}
%\begin{center}
\begin{figure}[ht]
\psfig{file=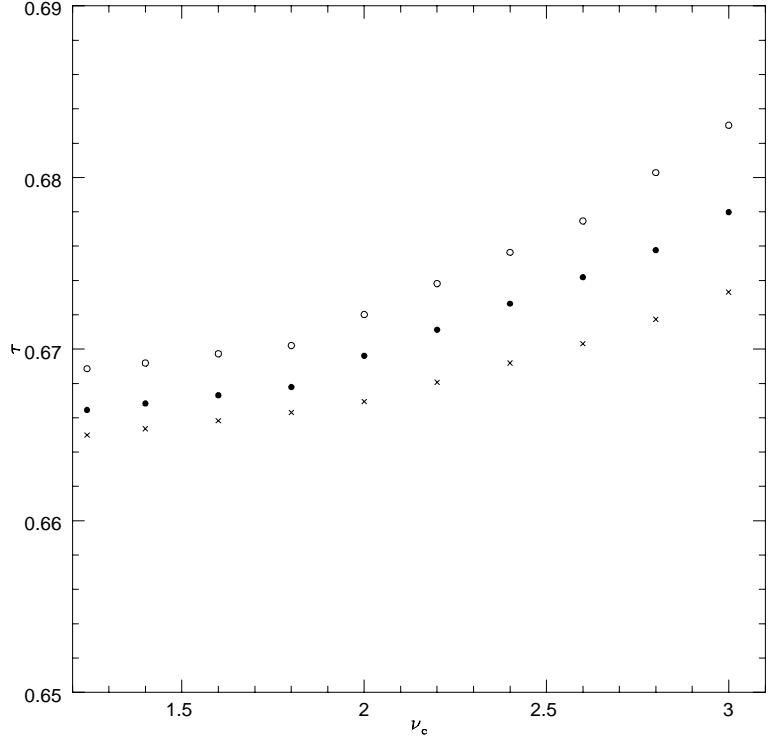,width=11cm}
\caption[h]{Collapse time $ \tau$
%Temporal evolution of the expansion parameter 
of a shell of matter made of galaxies and substructure 
when dynamical friction is taken into account,
versus $ \nu_{\it c}$
%unclustered substructure. The solid line is $ a(\tau)$  when 
%when it is taken into account.}
We assume a core radius of $r_{0} = 1  h^{-1} Mpc $ 
and a fixed correlation $ \Xi = 10^{3}$. 
{ \it Open circles:} $ R_{f} = 0.74 h^{-1} Mpc$; {\it filled circles:}
$ R_{f} = 0.65 h^{-1} Mpc$; {\it crosses:} $ R_{f} = 0.55 h^{-1} Mpc$.}
\end{figure}
%\end{center}
\newpage
After having determinated $ \eta$ solving numerically
eq. (\ref{eq:q}),  we get $ \tau$ as a function of 
$ \nu_{\it c}$,  $ R_{\it f}$,  $ r_{\it 0}$ and $ \Xi$
solving eq.(\ref{eq:p}). We perform these calculations 
for different set of values of $ \nu_{\it c}$,  $ R_{\it f}$,  $ r_{\it 0}$
and  $ \Xi$; the results are shown in 
figs. $1 {\scriptstyle \div} 4$.
Before commenting upon the figures, we want to remark that the dependence 
of $ \tau$ on $ \overline{\delta}$ is  qualitatively shown in fig. 5 
of {\sc AC94}. 
We observe that for $ \overline{\delta} > 10^{-2}$ 
the collapse time in presence of dynamical 
friction is always larger than in the imperturbated case but the 
magnitude of the deviation is negligible for larger $ \overline{\delta}$, 
whilst for $ \overline{\delta}  \leq 10^{-2}$ the deviations 
increase steeply with lower $ \overline{\delta}$. Then, having considered
$ \overline{\delta} = 0.01$, the estimation we get for $ \tau$  in  
\S 4 must be considered as a lower limit.
%\begin{center}
\begin{figure}[ht]
\psfig{file=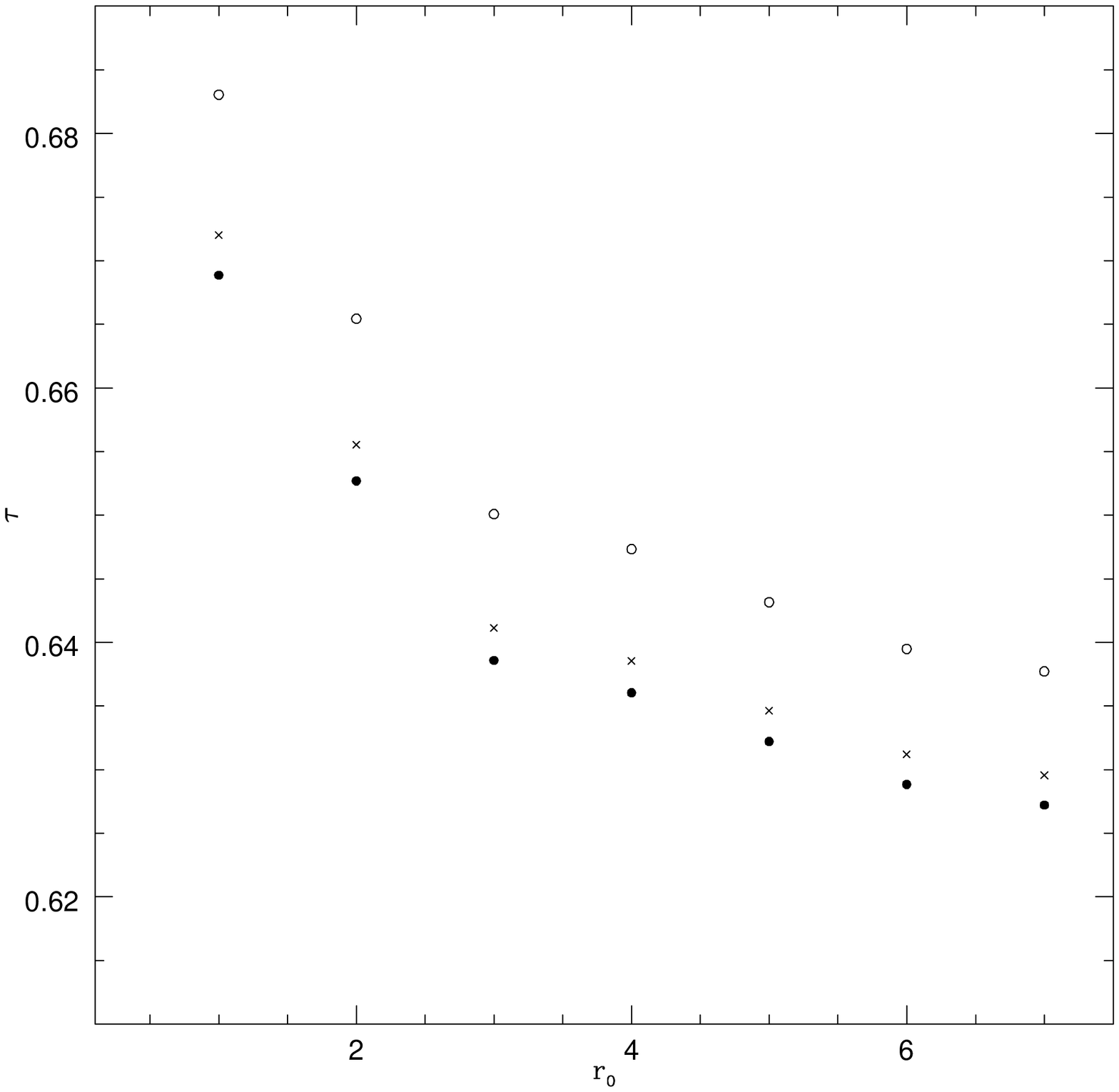,width=11cm}
\caption[h]{Collapse time $ \tau$
%Temporal evolution of the expansion parameter 
versus $ r_{0}$
We assume 
a filtering radius $ R_{f} = 0.74 h^{-1} Mpc$.
and a total number of peaks of substructure $ \Xi = 10^{3}$. 
{ \it Open circles:} $ \nu_{c} = 3 $; {\it crosses:} $ \nu_{c} = 2$; 
{\it filled circles:} $ \nu_{c} = 1.24$.}
\end{figure}
%\end{center}
In figure 1 we show the collapse time in presence of dynamical
friction, versus the peaks'  height, for different values of $ \Xi$. 
In this picture, we  show how $ \tau$ grows for larger values of $ \nu_{\it c}$
and for larger values of $ \Xi$. Similarly,  in fig. 2  we note 
how $ \tau$ increases for larger values of $ \nu_{\it c}$ and of $ R_{f}$.
The slope of the curves 
confirm our prevision on the 
behaviour of the collapse of a shell of baryonic matter falling into the 
central regions of a cluster of galaxies in the presence of dynamical friction:
the dynamical friction slows down the collapse (as {\sc DPG96} had already shown) 
and the effect, as we are showing in figs. 1 and 
2, increases as $\Xi$, $ R_{\it f}$, $\nu_{\it c}$ grow.
Here we want to remind that we are considering only the peaks of the local
density field with central height $ \nu$ larger than a critical 
threshold $ \nu_{\it c}$. This latter quantity is chosen to be the 
threshold at which $ r_{\it peak}$ ($ \nu \geq \nu_{\it c}$) $<<$ $ l_{\it 
av}$ where $ r_{\it peak}$ is the typical size of the peaks and $ l_{\it 
av}$ is the average peak separation (see also Bardeen et al. 1986).\\
\begin{figure}[ht]
\psfig{file=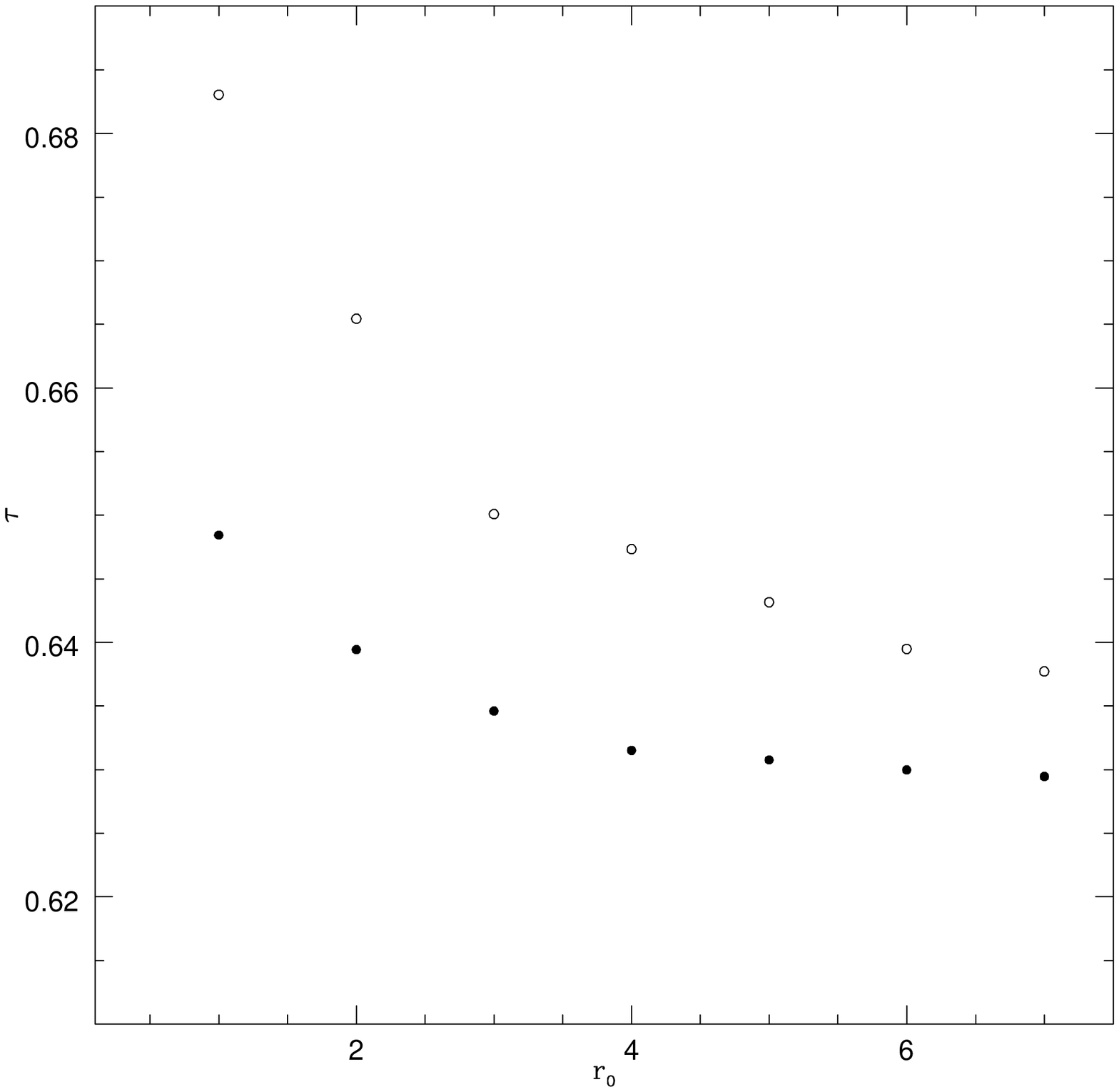,width=11cm}
\caption[h]{ Collapse time $ \tau$
of a shell of matter made of galaxies and substructure 
when dynamical friction is taken into account,
versus $ r_{0}$
We assume 
a filtering radius $ R_{f} = 0.74 h^{-1} Mpc$ 
and a peaks'  height $ \nu_{\it c} = 3$.
{\it Open circles:} $ \Xi = 10^{3}$; {\it filled circles:} $ \Xi 
= 10^{2}$.}
\end{figure}
In figs. 3 and 4 we show how the collapse time varies with the
core radius of the protocluster $ r_{0}$.
Note how  $ \tau$ grows  as $ r_{0}$ decreases: the smaller 
the core of the protocluster, the larger the time of collapse in the presence 
of dynamical friction; besides we show how this effect increases for 
larger values  of both  $ \nu_{\it c}$ and $ \Xi$.
%$ N_{\it tot}$.\\
~\\
~\\

%\begin{flushleft}
\section{ Conclusions and discussion}
%\end{flushleft}

In the first part of this work we have shown how the collapse time $ \tau$
of a shell of baryonic matter made of galaxies and substructure depends
on some parameters. In figures 1 and 2 we can see how $ \tau$ grows when
$ \Xi$ or $ R_{f}$ or $ \nu_{c}$ increases. It means that the effects of the 
presence of the dynamical friction should be more evident in the outer
regions of rich clusters of galaxies. Besides, we show how the collapse 
time of an infalling shell increases with decreasing values of $ r_{0}$, 
and becomes very large for $ r_{0} \leq 2  Mpc$ (see fig. 3). Then the 
slowing down of the collapse of an outer shell, of baryonic matter, 
within a cluster of galaxies owing to the dynamical friction is more 
remarkable in the clusters with core of little dimension.\\
~\\
~\\

%\newpage
%\begin{flushleft}
\acknowledgments
%\end{flushleft}

We are grateful to V. Antonuccio-Delogu for helpful and stimulating
discussions during the period in which this work was performed. 
~\\
~\\

%\newpage


\begin{references}
\reference  Antonuccio-Delogu, V., 1992, Ph.D. dissertation, ISAS, Trieste
\reference Antonuccio-Delogu, V., Atrio-Barandela, F., 1992, Ap.J 392, 403 (AA92) 
\reference Antonuccio-Delogu, V., Colafrancesco, S., 1994, Ap.J. 427, 72 (AC94)
\reference Bardeen, J.M, Bond, J.R., Kaiser, N., Szalay, A.S., 1986, 
Ap.J 304, 15
\reference Colafrancesco, S., Lucchin, F., Matarrese, S., 1989, Ap.J 345, 3
\reference Colafrancesco, S., Antonuccio, V., Del Popolo, A., 1995, Ap. J 
\reference Chandrasekhar, S., von Neumann, J., 1942, Ap.J 95, 489 (CvN)
\reference Chandrasekhar, S., von Neumann, J., 1943,Ap.J 97,1
\reference Del Popolo, A., Gambera, M.,  accepted for the pubblication
Astr. \& Astrop. (DPG96)
\reference Efstathiou, G., Rees, M.J., 1988, MNRAS, 230, 5p. 
\reference Gunn, J.E., Gott, J.R., 1972, Ap.J 176, 1 (GG72)
\reference Kaiser, N., 1993, in Cosmic velocity 
fields, ed. F.R. Bouchet, M. Lachieze-Rey, (Gif-sur-Yvette: Editions 
Frontieres), 533
\reference Kandrup, H.E., 1980, Phys. Rep. 63, n 1, 1
\reference Kolb, E.W., Turner, M.S., 1990, The early Universe (Addison-Wesley)
\reference Liddle, A.R., Lyth, D.H.,1993, Phys. Rep. 231, n 1, 2
\reference Peacock, J.A., Heavens, A.F., 1990, MNRAS 243, 133
\reference Ryden, B.S., 1988, Ap.J 329, 589
\reference Ryden, B.S., \&  Gunn, J.E., 1987, Ap.J 318, 15
\end{references}
\end{document}